\documentclass[11pt,paper,letterpaper]{JHEP3}
\usepackage{epsfig}
\usepackage{psfrag}
\usepackage{amsmath}
\newcommand{\bq}{\begin{eqnarray}}
\newcommand{\eq}{\end{eqnarray}}
\newcommand{\bqs}{\begin{eqnarray*}}
\newcommand{\eqs}{\end{eqnarray*}}

\def\a{\alpha}
\def\b{\beta}
\def\pd{\partial}

\def\d{\delta}
\def\e{\varepsilon}

\def\f{\phi}
\def\g{\gamma}
\def\k{\kappa}   
\def\l{\lambda}
\def\m{\mu}
\def\n{\nu}                
\def\th{\theta}
\def\r{\rho}                                   
\def\s{\sigma}
\def\z{\zeta}                
\def\O{\Omega}
\title{On stationary metrics in five dimensions}
\author{Pieter-Jan De Smet\\
C.~N. Yang Institute of Theoretical Physics\\
State University of New York\\
Stony Brook, NY 11794-3840, USA\\
E-mail: {\tt Pieterj@insti.physics.sunysb.edu} }
\preprint{YITP-SB-03-25 \\{\tt gr-qc/0306026}}
\abstract{It is well-known that the Kerr-metric 
(rotating black hole in four dimensions)
has Petrov type $D$. We prove a similar property in five dimensions. 
The Myers-Perry metric (rotating black hole in five dimensions) 
with one non-zero angular momentum has Petrov type 
\underline{22}. Conversely, we show that the Myers-Perry solution
is unique within a certain restricted class of metrics of Petrov 
type~\underline{22}.}
\begin{document}
\section{Introduction}
The metric of a four-dimensional rotating mass is described by the Kerr-metric.
This metric is algebraically special: it has Petrov type~D. 
As far as we know, there is no easy mathematical or
physical reason for this surprising result. 
The Schwarzschild metric also has Petrov type~D. 
However, this already follows from the isometries of the metric, without
using the detailed form of it. A rotating black hole has very few isometries
and it has taken physicists a long time to find its metric. Kerr found it
by looking within the class of metrics of Petrov type~D; he was lucky
to find it there.

In this paper we point out that something similar holds in five dimensions. The
five-dimensional Schwarzschild metric has Petrov type~\underline{22} (see 
ref~\cite{0206106} for a discussion of the five-dimensional Petrov classification). 
This is not surprising -- it follows again from the isometries. 
Things are more interesting for rotating black holes.
In Section~\ref{sec:MP}, 
we prove that the Myers-Perry solution~\cite{Myers-Perry} with one non-zero angular momentum
has Petrov type~\underline{22} as well. In Section~\ref{sec:class}, we prove
a converse statement: the Myers-Perry solution is the only 
five-dimensional metric satisfying the following conditions.
\begin{enumerate}
\item
The metric has 3 commuting Killing vectors and is invariant 
under 2 discrete $\mathbb{Z}_2$ transformations.
\item
The metric has Petrov type~\underline{22}.
\item
The metric is asymptotically flat at infinity.
\end{enumerate}
\section{The Petrov type of the Myers-Perry solution}\label{sec:MP}
In this section, we derive the Petrov type of the Myers-Perry 
metric~\cite{Myers-Perry}
\begin{equation}\label{sol:MP}
\begin{split}
ds^2 =& -\dfrac{R^2}{\r^2}\left(dt - a \sin^2\th d\f\right)^2 + 
    \dfrac{\sin^2\th}{ \r^2} \left(\left(r^2 + a^2\right) d\f - a dt\right)^2 + \\
& \dfrac{\r^2}{R^2} dr^2 + \r^2 d\th^2 + r^2 \cos^2\th d\psi^2,
\end{split}
\end{equation} with $\r^2 = r^2 + a^2 \cos^2\th$ and $R^2 = r^2 - 2 M + a^2$.
The Petrov classification in five dimensions~\cite{0206106} 
is basically an invariant classification of the 
Weyl spinor 
$$\psi_{abcd} = (\g^{ij})_{ab} (\g^{kl})_{cd} C_{ijkl},$$
where $C_{ijkl}$ is the Weyl tensor and $\g_i$ are the five-dimensional gamma 
matrices\footnote{We use the following convention for the gamma matrices:
$
\g_1 = i \s_1 \otimes 1,
\ \g_2 = \s_2 \otimes 1,
\ \g_3 = \s_3 \otimes \s_1,
\ \g_4 = \s_3 \otimes \s_2,
\ \g_5 = \s_3 \otimes \s_3\ .
$}.
We choose the following dual tetrad
\bqs
&& e^t = \dfrac{R}{\r} \left(dt - a \sin^2\th d\f\right), \qquad
e^\f =  \dfrac{\sin\th}{ \r} \left(\left(r^2 + a^2\right) d\f - a dt\right),\\
&& e^r =\dfrac{\r}{R} dr,\qquad
e^\th = \r d\th, \qquad e^\psi = r \cos\th d \psi.
\eqs
A calculation then gives the Weyl polynomial 
$$
\Psi = \psi_{abcd} x^a x^b x^c x^d= \dfrac{48 M}{\r^2}
\left( \dfrac{(x^1)^2}{z} - \dfrac{(x^2)^2}{\bar z}
-\dfrac{(x^3)^2}{\bar z}+ \dfrac{(x^4)^2}{z}\right)^2,\\
$$ 
with $z = r + i a \cos\th$. This polynomial is the square of a polynomial
of degree~2, hence, the Myers-Perry solution has Petrov type~\underline{22} in 
the notation of ref.~\cite{0206106}.
\section{The Myers-Perry solution is unique}\label{sec:class}
We now prove that the Myers-Perry solution is unique within a certain restricted
class of metrics of Petrov type~\underline{22}. The outline of the proof is as 
follows. We make ans\"atze for the tetrad, the connection $\omega$ and 
the Weyl spinor
$\Psi$ and solve the equations
\bq
&& \O =d\omega + \omega\wedge\omega,\label{eq:curv}\\
&& d\O + [\omega,\Omega]=0.\label{eq:Bianchi}
\eq
In these equations, the curvature $\Omega$ is completely determined by the Weyl
spinor $\Psi$ because the Ricci tensor vanishes. The condition on the Petrov type
simplifies the Weyl spinor $\Psi$ drastically. This makes it possible
to find all solutions. The ans\"atze can be found in section~\ref{s:ansatz},
the simplification of the Weyl spinor is discussed in section~\ref{s:Petrov}
and the solutions of the field equations are given in section~\ref{s:solutions}.
It can be seen that only the Myers-Perry solution is flat at infinity, which proves
its uniqueness. Although the Petrov condition simplifies the field equations
very much, the derivation of the solutions is still rather tedious. 
Some details can be found in appendix~\ref{app:ted}. 
\subsection{The ans\"atze}\label{s:ansatz}
Our coordinates are $t,\f, r,\th,\psi$. We assume that the metric has 
3 commuting Killing vectors which we choose as $\pd_t,\ \pd_\f$ and 
$\pd_\psi$. In addition, we assume that the metric is invariant
under two $\mathbb{Z}_2$ transformations:
\begin{itemize}
\item $(t,\f)\to(-t,-\f)$, this reflects that the metric is invariant if the rotation
reverses and time flows backwards
\item
$\psi \to -\psi$\ .
\end{itemize}
The most general ansatz for the tetrad is then as follows. 
The vectors $e_t$ and $e_\f$ are linear 
combinations of $\pd_t$ and $\pd_\f$, the vectors 
$e_r$ and $e_\th$ are linear combinations of 
$\pd_r$ and $\pd_\th$ and $e_\psi$ is proportional to $\pd_\psi$. The
coefficients in these linear combinations depend only on the variables 
$r$ and $\th$. The commutators then read
\begin{align}
[e_t,e_r] &= \mu e_t + \alpha e_\f   & [e_\f,e_r] &= \kappa e_t + \zeta e_\f\notag\\
[e_t,e_\th] &= \nu e_t + \beta e_\f  & [e_\f,e_\th] &= \lambda e_t + \eta e_\f\notag\\
[e_r,e_\psi] &= \sigma e_\psi        & [e_r,e_\th] &= \delta e_r + \e e_\th\label{comms}\\
[e_\th,e_\psi] &= \tau e_\psi\notag
\end{align}
It can be shown that the most general ansatz for the Weyl spinor consistent
with the above tetrad, is
\begin{equation}\label{ansatzpsi}
\begin{split}
\psi_{1111} &= \bar\psi_{3333}=
(T_{11} - T_{21}) + i (T_{12} - T_{22})\\
\psi_{1122}&=\bar\psi_{3344}= W_1 + i W_2\\
\psi_{1133} &= V_1 - V_2\\
\psi_{1144} &= \bar\psi_{2233}=I_2 + i I_3\\
\psi_{2222} &= \bar\psi_{4444}=
(T_{11} + T_{21}) + i(T_{12} + T_{22})\\
\psi_{2244} &= V_1 + V_2\\
\psi_{1234} &= \frac{1}{2} I_1
\end{split}
\end{equation}
The ansatz for the tetrad breaks the local Lorentz group $O(1,4)$ to 
$O(1,1)_{t\f} \times O(2)_{r\th}$. Under this group the ansatz for the Weyl spinor
transforms as
$$ \underbrace{{\bf (2,2) }}_{T_{ij}} \oplus
\underbrace{{\bf (2,1 )} }_{V_i} \oplus 
\underbrace{{\bf (1,2) }}_{W_i}\oplus 
\underbrace{2 \cdot {\bf (1,1) }}_{I_1,I_2}\oplus
\underbrace{{\bf (1',1') }}_{I_3}$$
\subsection{The condition on the Petrov type}\label{s:Petrov}
The Weyl polynomial associated with the Weyl spinor~\eqref{ansatzpsi}
is
$\Psi = \psi_{abcd}\ x^a x^b x^c x^d$.
It can be written as a square of a polynomial 
of degree~2 if and only if
one of the following four cases holds.
\begin{itemize}
\item[] Case A: $I_1 = 0$, $W_1^2+W_2^2=V_1^2-V_2^2=I_2^2+I_3^2$
and an extra condition on $T_{ij}$: 
\begin{itemize}
\item If $V_i$ is light-like, this extra condition is $T_{ij}=0$.
\item If $V_i$ is time-like, it is difficult to write this extra condition
in a manifestly $O(1,1)_{t\f} \times O(2)_{r\th}$ invariant way. If we choose
the particular frame $W_2=V_2=0$ and $W_2=V_1$, it reads
$T_{11} = 3 I_2,\ T_{22} = -3 I_3$ and $T_{12}=T_{21}=0$.
\end{itemize}\label{eq:PetrovA}
\item[]Case B: $I_2=I_3=T_{ij}=V_i=0$ and $W_1^2 + W_2^2 =  I_1^2$\ .
\item[]Case C: $I_2=I_3=T_{ij}=W_i=0$ and $V_1^2 - V_2^2 = I_1^2$\ .
\item[]Case D: $T_{ij}=W_i=V_i=0$ and $I_2^2 + I_3^2 = I_1^2$\ .
\end{itemize}
\subsection{Solutions of the field equations}\label{s:solutions}
The classification contains the following metrics
\begin{enumerate}
\item
The Kasner metric with exponents 
$(-1/2, 1/2,1/2,1/2)$
\begin{equation}\label{Kasner}
ds^2 = -dt^2 + \dfrac{1}{t} dx^2 + t (dy^2 +dz^2 + du^2)
\end{equation}
This metric is not asymptotically flat.
\item
\begin{equation}\label{metriek1}
\begin{split}
ds^2 =& -\frac{1}{r^2} \left( dt + 2 a  f(\th) d\f\right)^2 + 
    r^2 f'(\th)^2 d\f^2\\
    & + \left(\frac{b}{r^2} - \frac{a^2}{r^4} \right)^{-1}dr^2 + 
    r^2 d\th^2 + (b r^2 - a^2)d\psi^2,
\end{split}
\end{equation}
with $f(\th)$ a polynomial of degree~2. If one does a Wick transformation $r\to ir$,
the metric can be interpreted as a five-dimensional spatially homogeneous cosmological 
model\footnote{I would like to thank S.~Hervik for pointing this out.}~\cite{Hervik}. 
The isometry algebra is $A_{4,10} \oplus \mathbb{R}$ in the notation of 
ref.~\cite{algebras}. The metric reduces to the Kasner metric if $a =0$. This metric
is not asymptotically flat. 
\item
\begin{equation}\label{sol:gen}
\begin{split}
ds^2 =& -\dfrac{\tilde R^2}{\r^2}\left(dt - a \sin^2\th d\f\right)^2 + 
    \dfrac{k_2 - k_1 \cos^2\th}{ \r^2} \left(\left(r^2 + a^2\right) d\f - a dt\right)^2 + \\
& \dfrac{\r^2}{\tilde R^2} dr^2 + \dfrac{\sin^2\th}{k_2 - k_1 \cos^2\th }\r^2 d\th^2 
+ r^2 \cos^2\th d\psi^2,
\end{split}
\end{equation} where $\r^2 = r^2 + a^2 \cos^2\th$ and 
$\tilde R^2 = k_1 r^2 - 2 M + k_2 a^2$.
This metric has Lorentz signature only if $k_2\ge k_1\ge0$. If $k_1\neq 0$, the metric
is actually the Myers-Perry solution~\eqref{sol:MP}. 
If $k_1=0$, it is not asymptotically
flat.
\end{enumerate}
Hence, we see that the Myers-Perry solution is the unique metric that
(i) has three commuting Killing vectors $\pd_t$, $\pd_\f$ and $\pd_\psi$, 
(ii) is invariant under $(t,\f)\to (-t,-\f)$ and $\psi\to-\psi$,
(iii) is asymptotically flat and 
(iv) has Petrov type~\underline{22}.
\section{Conclusions}\label{s:conc}
The well-known ``no-hair'' theorems
of four-dimensional gravity do not hold in five dimensions for the general
stationary case. For example, in~\cite{ring} a rotating ring was found
with the same mass and angular momentum as the Myers-Perry solution, but
with a different horizon topology. In the static case on the other hand, 
one can prove that asymptotically flat solutions are unique~\cite{Gibbons,Hwang}.
See also ref.~\cite{Kol} for a speculation that uniqueness properties hold in 
higher dimensions if one (i) specifies the horizon topology and (ii) only considers
stable solutions. The uniqueness theorems can also be saved by adding supersymmetry:
in ref.~\cite{Reall} it is proven that the BMPV solution~\cite{BMPV} is the only 
supersymmetric black hole of minimal $N=1$, $D=5$ supergravity. 
It is nice that the Myers-Perry solution is unique within
the class of metrics of Petrov type~\underline{22}. This result could shed some light 
on the structure of solutions of five-dimensional gravity.

It would be interesting to generalize the above analysis. Some possibilities are the 
following.
\begin{itemize}
\item
We have restricted our study of the Myers-Perry solution by putting one of the 
angular momenta to zero. The above analysis can probably be generalized to the 
case with two angular momenta.
\item
Preliminary investigations indicate that the rotating ring found by Reall~\cite{ring} 
has Petrov type~22.
The generalization of this rotating ring to two non-zero angular momenta is not
known. It is possible that it can be found in the class of metrics of Petrov 
type~22.
\item 
The charged generalization of the Myers-Perry solutions is not known. The following
reasons suggest that it might be of Petrov type~\underline{22}.
\begin{itemize}
\item
As shown in this paper, the uncharged limit has Petrov type~\underline{22}.
\item
The non-rotating limit (five-dimensional Reissner-Nordstr\o m metric) has
Petrov type~\underline{22} as well.
\item
The charged rotating black hole in four dimensions (the Kerr-Newmann metric) has 
the same Petrov type as the uncharged rotating black hole, i.e.~Petrov type~$D$.
\end{itemize}
It seems worthwhile to look for the charged generalization 
within the class of metrics of Petrov type~\underline{22}.
\end{itemize}
\section*{Acknowledgments}
This work has been supported in part by the NSF grant PHY-0098527.
It is a pleasure to thank H.~Reall, S.~Hervik, G.~Gibbons and W.~Siegel 
for discussions.
\appendix
\section{More details}\label{app:ted}
In this appendix, we solve the Einstein equations with the 
ans\"atze~\eqref{comms} and~\eqref{ansatzpsi}. As discussed in section~\ref{s:Petrov},
the Petrov condition leads to four different cases. These are discussed separately.
\subsection*{Case A}
We only discuss the case where $V_i$ is time-like. $V_i$ light-like is a 
special case of Case~C, which is discussed below.
We assume $V_1\neq0$, otherwise, it follows from~\eqref{eq:PetrovA} that
the space is flat. From the Bianchi equations~\eqref{eq:Bianchi}, 
we obtain $\k=\b=0$, $\e=-\z$, $\d=\n$ and \\
\begin{minipage}{7.5cm}
\bqs
&&\tau I_3= \a V_1\\
&&\tau I_2 = -(2 \n +\tau) V_1
\eqs
\end{minipage}
\begin{minipage}{7.5cm}
\bqs
&&\s I_3 = \l V_1\\
&&\s I_2 = (2 \z + \s) V_1
\eqs
\end{minipage}
\\[2ex]
Combining the above equations with the Petrov condition~\eqref{eq:PetrovA} leads to
\bq
&&\n \s + \tau \s + \tau \z =0\label{eq:a}\\
&& \a^2 + 4 \n^2 + 4  \n \tau=0\label{eq:d1}\\
&&\l^2 + 4 \z^2 + 4 \z \s=0\label{eq:d2}
\eq
Furthermore, combining~\eqref{eq:a} with~\eqref{eq:curv} gives
\begin{equation}\label{eq:f}
\l\a = 4 \z\n
\end{equation}
At this point, we want to choose coordinates such that $e_r = A \z \pd_r$ and 
$e_\th = B \n \pd_\th$, where $A$ and $B$ are two functions. However, we can
only make this choice if $\z$ and $\n$ are both non-zero. Therefore, we have to
treat the cases $\z=0$ or $\n=0$ separately. This is a bit unfortunate because 
it will turn out that the solutions in the latter cases can be obtained as limits
of solutions
in the case $\z\neq0\neq\n$. Anyway, 
we break up case A into 5 subcases. 
\subsubsection*{Case A.1: $\n\neq0$ and $\z\neq0$}
From~\eqref{eq:curv} and~\eqref{eq:f}, one obtains 
$e_r\n = - 3 \z \n$ and $e_\th\z = - 3 \z\n$.
Using these equations, one can show that there are coordinates such that
$$e_r = \frac{\r^2}{r} \z \pd_r \qquad\text{and}\qquad 
e_\th = - \frac{\r^2}{a^2 \sin\th \cos\th} \n \pd_\th,$$
where $\r^2 = r^2+a^2\cos^2\th$. The solution of~\eqref{eq:curv} and~\eqref{eq:Bianchi}
is the metric~\eqref{sol:gen}.
\subsubsection*{Case A.2: $\n=0$ and $\tau=0$}
From the Bianchi identities, one obtains $\a =0$ and $\m=-\z$. The resulting
field equations are then easily solved. The solution is the metric~\eqref{metriek1}.
\subsubsection*{Case A.3: $\n=0$ and $\tau\neq0$}
From the Bianchi identities, one obtains $\l=\a=0$, $I_3=0$, 
$I_2=-V_1$ and $\z=-\s$. The solutions in this
case are the Schwarzschild metric and the Kasner metric~\eqref{Kasner}.
\subsubsection*{Case A.4: $\z=0$ and $\s=0$}
This case leads again to the metric~\eqref{metriek1}.
\subsubsection*{Case A.5: $\z=0$ and $\s\neq0$}  
This case leads again to the Schwarzschild metric or the Kasner metric.
\subsection*{Case B}
We can choose $W_2=0$ and $W_1=I_1$ by  
an $O(2)_{r\th}$ transformation. Furthermore, we can use an $O(1,1)$ boost to put $\a=\b=0$.
The Bianchi equations yield $\k=0$, $\e=-\z$, $\m=\z$, $\d=0$ and $\tau=0$. Using the 
commutation relations~\eqref{comms}, we can choose coordinates in which the metric has the form
$$ds^2 = A(r) d\psi^2 + B(r) dr^2 + r^2 d\tilde s^2 (t,\th,\f).$$
From Einstein's equations, it directly follows that the three-dimensional metric 
$ d\tilde s^2$ is Einstein, hence has constant curvature. If $d\tilde s^2$ is Euclidean,
we obtain the Kasner metric~\eqref{Kasner}. If $d\tilde s^2$ is a sphere, we obtain the 
Schwarzschild metric.
\subsection*{Case C}
If $I_1=0$, we can choose $V_1=V_2$  by an $O(1,1)_{t\f}$ transformation. 
If $I_1\neq 0$, we make $V_2=0$ and $V_1=I_1$ by a boost. In both cases, Einstein's
equations lead to flat space.
\subsection*{Case D}
We put $\a=\b=\n=0$ by an $O(1,1)_{t\f} \times O(2)_{r\th}$ transformation. 
From the Bianchi identities, one can rather easily see that there are only solutions
if $I_3=0$ and $I_2=I_1$.
The Bianchi identities then give $\eta=\k=\l=0$ and $\z=\m=-\s$. 
The solution is the Kasner metric~\eqref{Kasner}.

\end{document}